\documentclass[12pt]{iopart}
 \usepackage{epsfig}

\def\gsim{\,\raise0.3ex\hbox{$>$\kern-0.75em\raise-1.1ex\hbox{$\sim$}}\,}
\def\lsim{\,\raise0.3ex\hbox{$<$\kern-0.75em\raise-1.1ex\hbox{$\sim$}}\,}

\usepackage{iopams}  
\usepackage{pslatex}
\usepackage{txfonts}

\newcommand{\sqrtsnn}{\sqrt{s_{_{NN}}}}

\providecommand{\mean}[1]{\ensuremath{\left<#1\right>}}

\begin{document}

\title{Direct photon spectra in Pb-Pb at $\sqrtsnn$ = 5.5 TeV: hydrodynamics+pQCD predictions}
\vspace{-2mm}
 
\author{Fran\c{c}ois Arleo$\hspace{1mm}^{1,2}$, David d'Enterria$\hspace{1mm}^2$, Dmitri Peressounko$\hspace{1mm}^3$}
\address{{\bf 1.} LAPTH, UMNR CNRS/Univ. Savoie, B.P. 110, 74941 Annecy-le-Vieux Cedex}%, France}%, Switzerland}
\address{{\bf 2.} CERN/PH, CH-1211 Geneva 23}%, Switzerland}
%\address{{\bf 1.} PH-TH, CERN, CH-1211 Geneva 23}%, Switzerland}
%\address{{\bf 2.} PH-PH, CERN, CH-1211 Geneva 23}%, Switzerland}
\address{{\bf 3.} RRC ``Kurchatov Institute'', Kurchatov Sq. 1, Moscow 123182}

\begin{abstract}
  The $p_T$-differential spectra for direct photons produced in Pb-Pb collisions at the LHC, 
  including thermal (hydrodynamics) and prompt (pQCD) emissions are presented.
\end{abstract}
%\maketitle

\noindent
We present predictions for the transverse momentum distributions of direct-$\gamma$ (i.e. photons 
not coming from hadron decays) produced at mid-rapidity in Pb-Pb collisions at $\sqrtsnn$ = 5.5 TeV based 
on a combined hydrodynamics+pQCD approach. Thermal photon emission in Pb-Pb at the LHC is computed 
with a hydrodynamical model successfully used in nucleus-nucleus collisions at RHIC energies~\cite{d'Enterria:2005vz}. 
The initial entropy density of the produced system at LHC is obtained by extrapolating empirically 
the hadron multiplicities measured at RHIC~\cite{dde_peress_arleo}. Above 
$p_T\approx$~3~GeV/$c$, additional prompt-$\gamma$ production from parton-parton scatterings 
is computed perturbatively at next-to-leading-order (NLO) accuracy~\cite{Aurenche:1998gv}.
We use recent parton distribution functions (PDF)~\cite{Pumplin:2002vw} and parton-to-photon 
fragmentation functions (FF)~\cite{Bourhis:2000gs}, modified resp. to account for initial-state 
shadowing+isospin effects~\cite{deFlorian:2003qf} and final-state parton energy loss~\cite{Arleo:2007}.\\
\vspace{-2mm}

\noindent
%We use the same cylindrically-symmetric boost-invariant 2+1-D relativistic hydrodynamics ofour previous works. 
We follow the evolution of the hot and dense system produced in central Pb-Pb at LHC
by solving the equations of (ideal) relativistic 2D+1 hydrodynamics~\cite{d'Enterria:2005vz,dde_peress_arleo} 
starting at a time $\tau_0 = 1/Q_s\approx$ 0.1 fm/$c$. The system is assumed to have an 
initial entropy density of $s_0$~=~1120~fm$^{-3}$, which corresponds to a maximum temperature 
at the center of $T_0\approx$ 650~MeV ($\mean{T_0}\approx$ 470 MeV). 
We use a quark gluon plasma (QGP) and hadron resonance gas (HRG) equation of state
above and below $T_{\rm crit}\approx$ 170 MeV resp., connected by a 
standard Maxwell construction assuming a first-order phase transition at $T_{\rm crit}$. 
%we chemically freeze-out the system (i.e. fix the hadron ratios) at $T_{\rm crit}$.
%explicitly conserving particle numbers by introducing individual (temperature-dependent)
%chemical potentials for each hadron.
Thermal photon emission is computed using the most recent parametrizations 
of the QGP and HRG $\gamma$ rates. For the QGP phase we use the AMY complete leading-log 
emission rates including LPM suppression~\cite{arnold}. For the HRG phase, we employ the
improved parametrization from Turbide {\it et al.}~\cite{turbide}.\\
\vspace{-2mm}

\noindent
Our NLO pQCD predictions are obtained with the code of ref.~\cite{Aurenche:1998gv}
with all %factorization and fragmentation 
scales set to $\mu=p_{T}$. Pb-Pb yields are obtained scaling the NLO cross-sections 
by the number of incoherent nucleon-nucleon collisions: $N_{\rm coll}$ = 1670, 12.9 for 0-10\% central 
($\mean{b}$ = 3.2 fm) and 60-90\% peripheral ($\mean{b}$ = 13 fm). 
Nuclear (isospin and shadowing) corrections of the CTEQ6.5M PDFs~\cite{Pumplin:2002vw} 
are introduced using the NLO nDSg parametrization~\cite{deFlorian:2003qf}. At relatively low 
$p_T$, prompt photon yields have a large contribution from jet fragmentation processes.
%[The typical LO quark-gluon Compton and $q-\bar{q}$ annihilation photon processes are subdominant
%below $p_T\approx$ 20(?) GeV/$c$]. 
As a result, final-state parton energy loss in central Pb-Pb affects also the expected prompt 
$\gamma$ yields. We account for medium-effects on the $\gamma$-fragmentation component 
by modifying the BFG parton-to-photon FFs~\cite{Bourhis:2000gs} with 
BDMPS quenching weights. The effects of the energy loss are
encoded in a single parameter, $\omega_c=\mean{\hat{q}}\,L^2\approx$ 50 GeV,
extrapolated from RHIC. The combination of initial-state (shadowing) and final-state 
(energy loss) effects results in a quenching factor for prompt photons %at the LHC 
of $R_{PbPb}\approx$ 0.2 (0.8) at $p_T$ = 10 (100) GeV/$c$~\cite{Arleo:2007}.\\
\vspace{-2mm}

\begin{figure}[htbp]
\begin{centering}
\hspace{6mm}
\includegraphics[width=0.45\textwidth,height=8.cm]{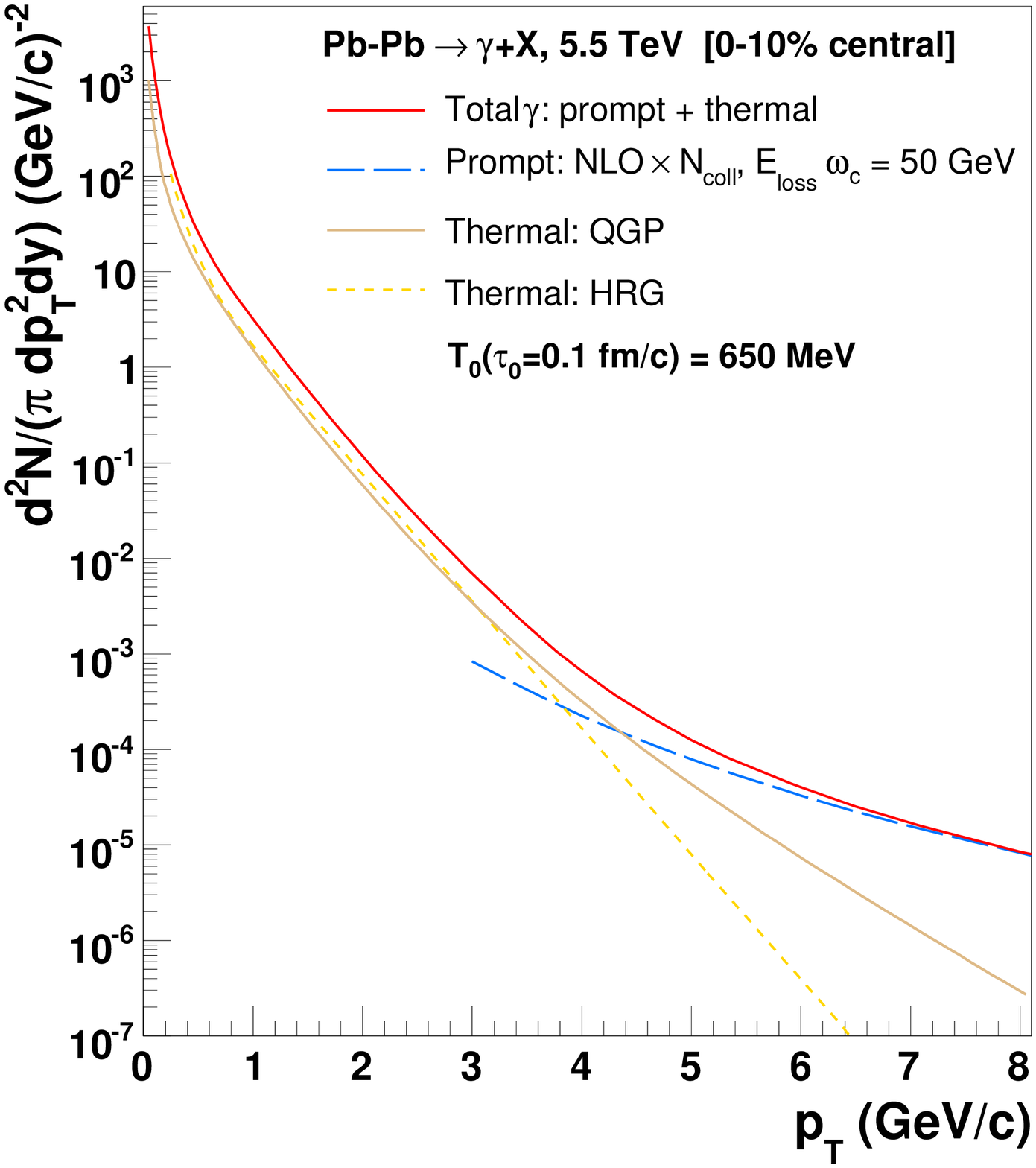}
\hspace{4mm}
\includegraphics[width=0.45\textwidth,height=8.cm]{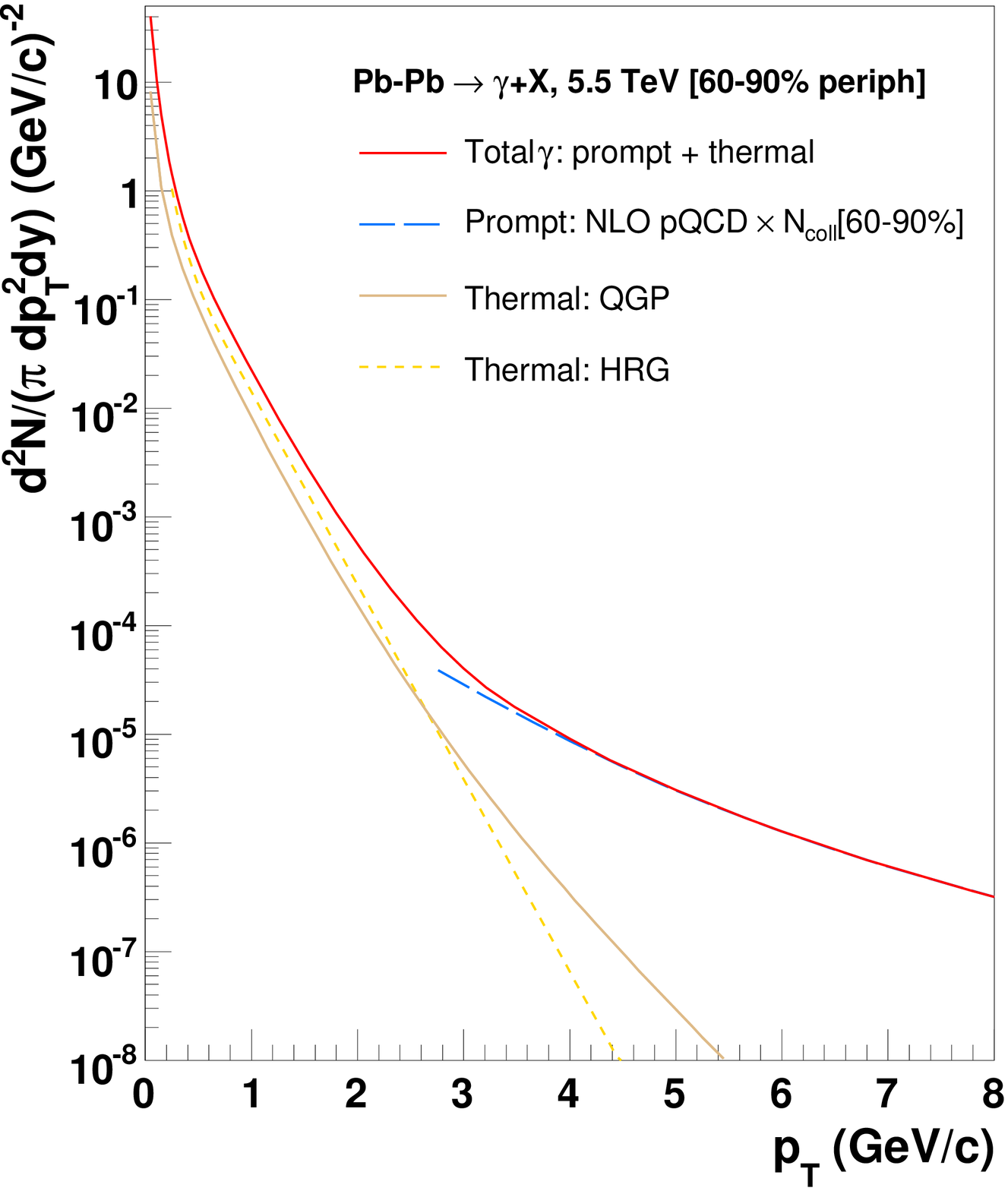}
\end{centering}
\vspace{-5mm}
\caption{\label{fig:spectra}Direct-$\gamma$ spectra in 0-10\% central (left) and 60-90\% peripheral (right) Pb-Pb at 
$\sqrtsnn$ = 5.5 TeV, with the thermal (QGP and HRG) and prompt (pQCD) contributions differentiated.}
\end{figure}

\noindent
Our predictions for the direct photon spectra at y=0 in Pb-Pb at 5.5 TeV are 
shown in Fig.~\ref{fig:spectra}. The thermal contribution dominates over the (quenched) 
pQCD one up to $p_T\approx$ 4 (1.5) GeV/$c$ in central (peripheral) Pb-Pb. Two differences
are worth noting compared to RHIC results~\cite{d'Enterria:2005vz}: 
(i) the thermal-prompt crossing point moves up from $p_T\approx$ 2.5 GeV/$c$ to 
$p_T\approx$~4.5~GeV/$c$, and (ii) most of the thermal production in this transition
region comes solely from the QGP phase. Both characteristics make of semi-hard 
direct photons at LHC, a valuable probe of the thermodynamical properties of the system.

%%%%%%%%%%%%%%%%%%%%%%%%%%%%%%%%%%%%%%%%%%%%%%%%%%%%%%%%%%%%%%%
\vspace{-2mm}
\vspace{-3mm}
\noindent
Dd'E. and D.P. acknowledge resp. support from 6th EU FP contract MEIF-CT-2005-025073 and
MPN Russian Federation grant NS-1885.2003.2.

%%%%%%%%%%%%%%%%%%%%%%%%%%%%%%%%%%%%%%%%%%%%%%%%%%%%%%%%%%%%%%%

\end{document}